\documentclass[twocolumn, 10pt]{article}

\usepackage[utf8]{inputenc}
\usepackage[english]{babel}
\usepackage[margin=1.8cm]{geometry}
\usepackage{authblk} 
\usepackage{amsmath, amssymb, amsfonts}
\usepackage{graphicx}
\usepackage{hyperref}
\usepackage{cite}
\usepackage{booktabs}
\usepackage{abstract}

\hypersetup{
    colorlinks=true,
    linkcolor=blue,
    urlcolor=cyan,
    citecolor=blue,
}

\title{\textbf{Optical Manipulation of Erythrocytes via Evanescent Waves: Assessing Glucose-Induced Mobility Variations}}

\author[1,2]{T. Troncoso Enríquez}
\author[1]{J. Staforelli-Vivanco\thanks{Corresponding author: jstaforelli@udec.cl}}
\author[3]{I. Bordeu}
\author[4]{M. González-Ortiz}
\affil[1]{\small Departamento de Física, Facultad de Ciencias Físicas y Matemáticas, Universidad de Concepción, Chile.}
\affil[2]{\small Departamento de Ingeniería Eléctrica, Universidad de Concepción, Chile.}
\affil[3]{\small Departamento de Física, Facultad de Ciencias Físicas y Matemáticas, Universidad de Chile, Chile.}
\affil[4]{\small Laboratorio de Investigación Materno-Fetal (LIMaF), Departamento de Obstetricia y Ginecología, Facultad de Medicina, Universidad de Concepción, Concepción, Chile}
\date{\today}

\begin{document}

\maketitle

%\twocolumn
%\begin{center}

\begin{abstract}
This study investigates the dynamics of red blood cells (RBCs) under the influence of evanescent waves generated by total internal reflection (TIR). Using a 1064 nm laser system and a dual-chamber prism setup, we quantified the mobility of erythrocytes in different glucose environments. Our methodology integrates automated tracking via TrackMate\textsuperscript{\copyright} to analyze over 60 trajectory sets. The results reveal a significant decrease in mean velocity, from 11.8 $\mu$m/s in 5 mM glucose to 8.8 $\mu$m/s in 50 mM glucose ($p=0.019$). These findings suggest that evanescent waves can serve as a non-invasive tool to probe the mechanical properties of cell membranes influenced by biochemical changes.
%\vspace{0.5cm}
\end{abstract}
%\end{center}
\maketitle

\section{Introduction}

The manipulation of micro-scale objects using light has revolutionized biophysics and soft matter research since the pioneering work of Arthur Ashkin on optical tweezers \cite{Petra2018,Ashkin2000,Kishan2006,Volpe2017,Mark2015}. While traditional optical trapping relies on the gradient forces of a focused Gaussian beam, the use of surface-bound electromagnetic fields, specifically evanescent waves, offers a unique paradigm for large-scale, non-invasive particle transport. Evanescent waves are generated via Total Internal Reflection (TIR) at an interface between two media of different refractive indices. These waves do not propagate into the secondary medium but instead decay exponentially with distance from the interface, typically within a few hundred nanometers \cite{Milan2013,Oleg2020,Derek2001,Griffiths2017}. This localized energy concentration allows for the application of radiation pressure to particles and cells strictly near the surface, minimizing bulk heating and photo-damage.

In recent years, the interaction between evanescent fields and biological entities has gained traction due to its sensitivity to surface properties. Red blood cells (RBCs), or erythrocytes, are of particular interest as they serve as primary indicators of rheological and biochemical changes in the human body \cite{Taitt2016,Oleg2019,Oleg2019b}. The mechanical deformability and mobility of RBCs are known to be affected by their environment, including osmolarity and glucose concentration. Chronic exposure to high glucose levels, a hallmark of conditions such as diabetes mellitus, can lead to glycation of membrane proteins and alterations in the lipid bilayer, ultimately increasing membrane rigidity and affecting the cell's hydrodynamic behavior.

Despite the extensive literature on optical trapping of RBCs, studies focusing on their lateral propulsion via evanescent waves remain limited. Most existing models, such as the one developed by Almaas and Brevik \cite{Almaas1995}, provide a robust framework for predicting forces on rigid spheres, yet the application to complex, deformable biological cells requires further experimental validation. 

In this work, we present a systematic study of the mobility of human erythrocytes under the influence of an evanescent field generated by a 1064 nm laser. By implementing a dual-chamber experimental setup, we compare the dynamics of RBCs in physiological (5 mM) and hyperglycemic (50 mM) glucose environments \cite{Tapia2021,Bravo2023}. We utilize automated tracking algorithms to quantify mean velocities and diffusion patterns, aiming to establish a correlation between biochemical environment and optical response. Our results provide evidence that evanescent wave propulsion can act as a sensitive, non-contact probe for detecting subtle changes in cellular biomechanics induced by glucose.

\section{Theoretical Background}

The evanescent field dynamics appears when a laser beam strikes a glass-liquid interface at an angle $\theta > \theta_c$.. The intensity decays as $I(z) = I_0 e^{-z/d_p}$, where $d_p$ is the penetration depth \cite{Mario2017}. For our system ($n_1=1.51$, $n_2=1.33$), $d_p$ is approximately 250 nm. The net force on a cell near the interface is a balance between the gradient force, the scattering force (radiation pressure), and the viscous drag $F_d = 6\pi\eta R v$. By measuring the steady-state velocity $v$, we can infer the optical force exerted by the EW \cite{Nieto2011,Griffiths2017}.

\section{Materials and Methods}

\begin{figure}[ht]
    \centering
\includegraphics[width=\linewidth]{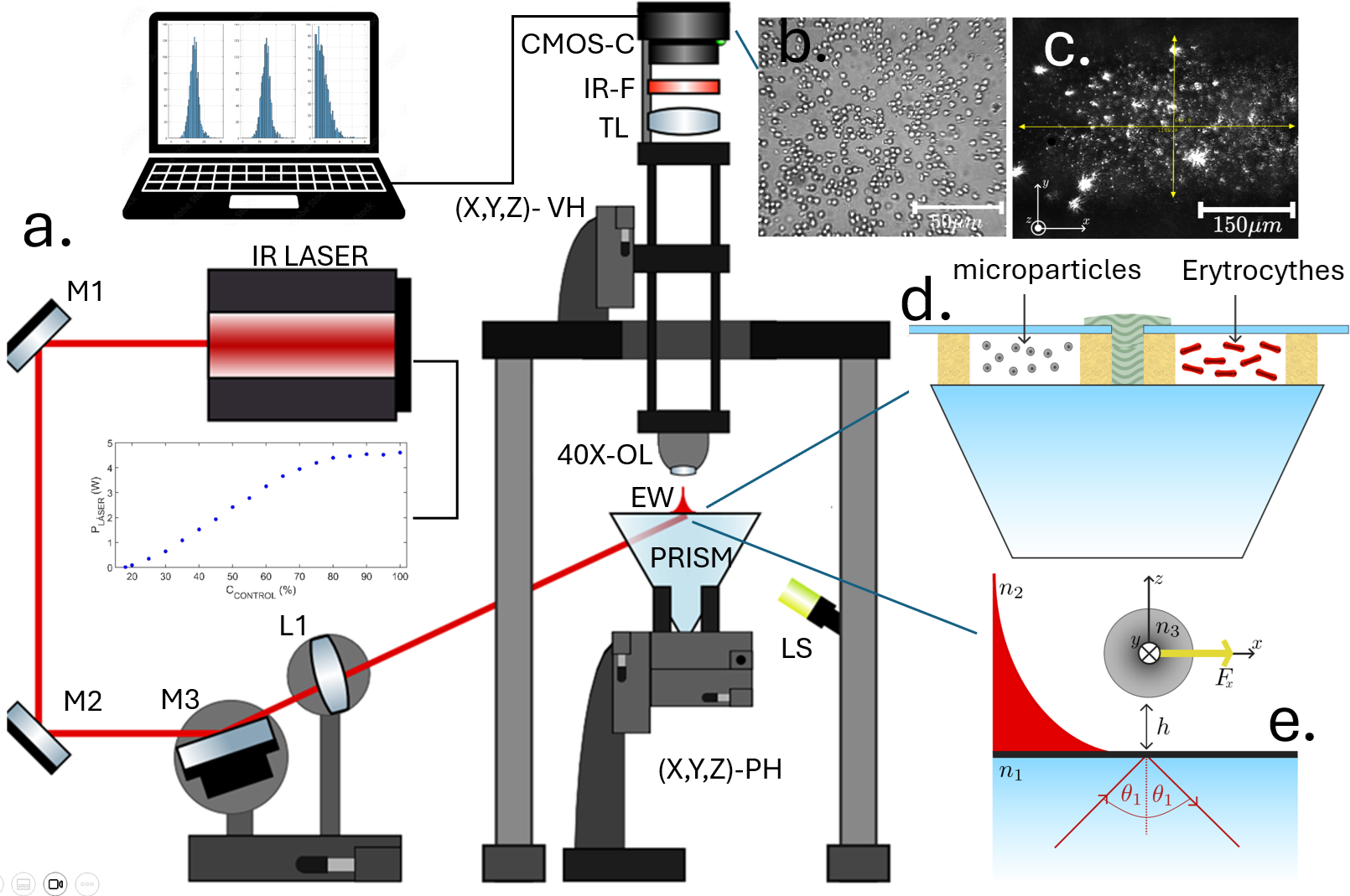}
\caption{The experimental setup consists of a. laser source 1064 nm CW laser (Ventus) operated at 1.8 W. A series of prisms fabricated with dual-region chamber to allow simultaneous measurement of control (polystyrene) and RBC samples. Imaging with 60x objective (NA 0.85) coupled with a CMOS camera at 25 fps. b. Red blood cell visualization. c. Major and minus axis of EW-area on prism surface. d. Two-chamber method on a single prism for simultaneous experiment with polystyrene micro-particles and erythrocytes (RBC). This method avoid misalignment and artifacts instead of single samples on different prisms.e. EW-scheme: a micro-particle of radius $a$ and refractive index $n_3$ located at a height $h$ above the interface, immersed in the evanescent field (EW) generated by a laser with an angle of incidence $\theta_1$.}
\label{fig1_setup}
\end{figure}

Experimental setup is shown in figure \ref{fig1_setup}. The system is mounted on an air-suspended optical table (1.5 × 2 m, Thorlabs) for maximum isolation. Main opto-mechanical components are: IR infrared laser source (Ventus 1064, Laser Quantum) connected to a control unit. The unit reach up to $5W$ output power at $90\%$ of seated current. Laser alignment is performed with a pair of reflective mirrors $M_1$ and $M_2$ $BB1-E03$ from Thorlabs\textsuperscript{\copyright}. $M_3$ is an elevation mirror. The tilt value $M_3$ is chosen so that the normal beam angle is as close as possible to the critical angle. $L_1$ $N-BK7$ biconvex lens $f= 400$ mm (LB1391, Thorlabs) focused the laser on surface prism. $(X,Y,Z)-PH$ is a three-axis controller position for prism holder and sample fabricated with an Ender v3 3D printer. EW is the evanescent wave generated in the sample at the interface with the prism. $L_S$ is a tungsten light source guided with fiber cord. $(X,Y,Z)-VH$ is a three-axis controller position of the vision holder system which is compound of a 60X objective lens from Edmund Optics, a bi-convex 10 cm tube lens, an IR- Filter and CMOS Monochrome sensor 1280 x 1024 DCC1545M from Thorlabs\textsuperscript{\copyright}, placed at $L \approx 250$ mm from Tube Lens to achieve the desired magnification and resolution. The CMOS-camera is connected to a desktop unit for data acquisition using Trackmate\textsuperscript{\copyright} \cite{Trackmate2017}, processing and analysis 

\section{Sample Preparation}

Adult blood samples (10 $\mu$l) were obtained from healthy volunteers and suspended in PBS solutions. The blood samples were diluted (4/1000) in phosphate-buffered saline (PBS) solution [mM: 130 NaCl, 2.7 KCl, 0.8 Na2HPO4, 1.4 KH2PO4 (pH 7.4)]. Then two groups were prepared: a physiological control (5 mM glucose) and a high-concentration group (50 mM glucose). These values were chosen to amplify the effects of glucose on the red blood cells. Glucose level is higher than normal for values$\>$6.1 mmol/L (110–125 mg/dL) and hyperglycemia for values higher than 7.0 mmol/L (126 mg/dL) \cite{Cho2018}. Polystyrene spheres ($D=1.02 \mu$m) served as the validation control.

The systematic and random errors identified during the development of the setup are crucial for interpreting the variance observed between different measurement sessions. The most significant source of systematic error was found during the replacement of prisms between samples. Each exchange required a manual realignment of the laser focus and the recording region. This process introduced variations in the local intensity of the evanescent field, directly affecting the optical force exerted on the cells. To avoid the systematic error, a configuration of double-chamber was designed to allocate RBC and micro-polystyrene control samples simultaneously on each prism surface.

\begin{figure}[ht]
    \centering
       \includegraphics[width=\linewidth]{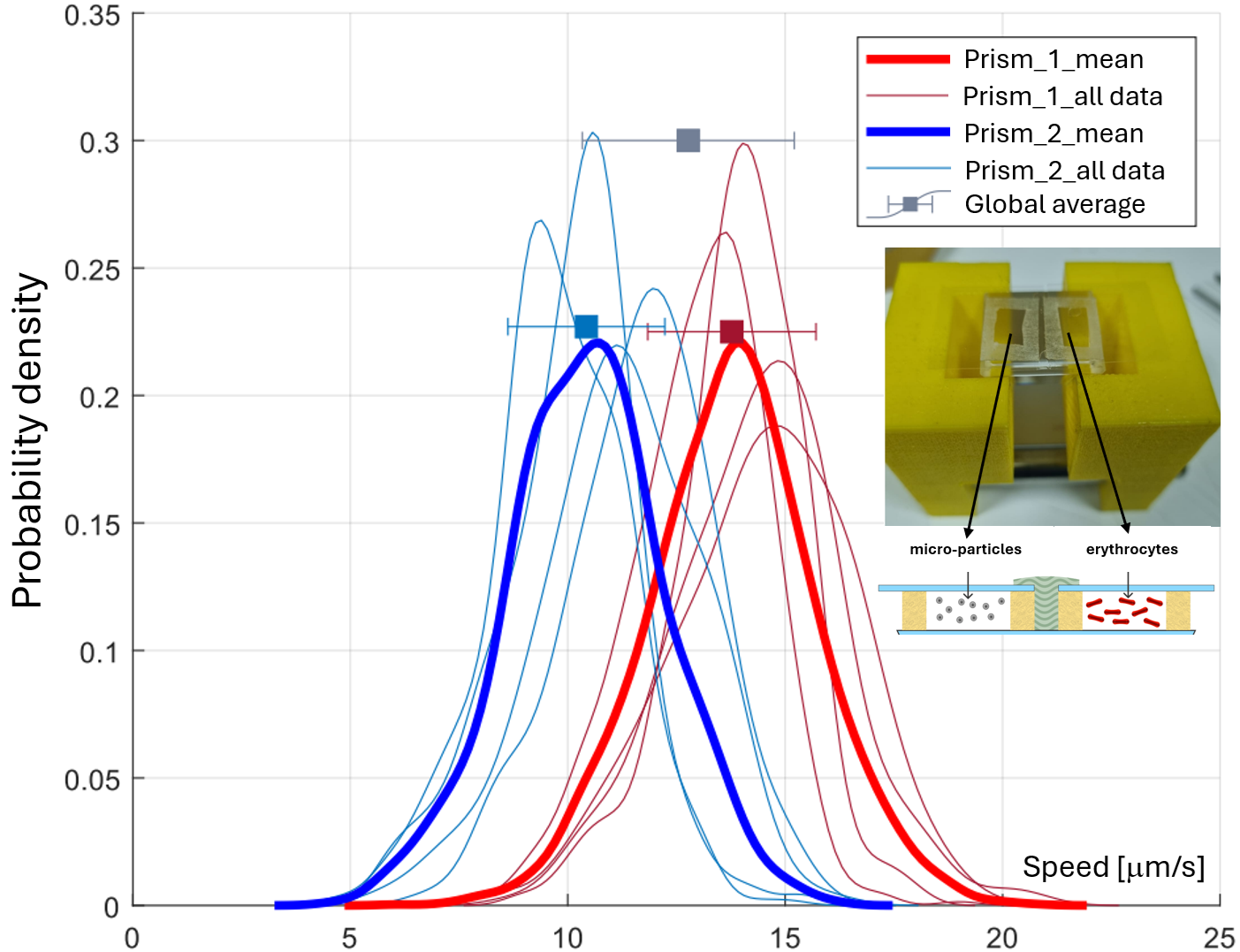}
    \caption{Probability densities for the velocity of microparticles, measured for samples on two different prisms.}
    \label{fig2_chambers}
    \end{figure}
    
From Figure \ref{fig2_chambers} is clear that the distributions obtained for one prism are significantly different from those obtained for the other, with an average velocity of 10.54 $\mu$m/s for prism 1 and 13.77$\mu$m/s for prism 2. Therefore, while the velocities recorded within the same prism are similar, the velocities measured between prisms appear to vary much more significantly from one another. A divided chamber on each surface prism permits to measure velocities of polystyrene control micro-spheres and erythrocytes simultaneously in the same sample prism.

\section{Theoretical model}

The comparison utilizes the model proposed by \textit{Almaas and Brevik} \cite{Almaas1995}, which calculates the optical force ($F_{opt}$) exerted by an evanescent field on a spherical particle. To translate this force into a theoretical velocity ($v_{th}$), the Stokes' Drag Law is applied:

\begin{equation}
    F_{opt} = 6\pi \eta r v_{th} \beta
\end{equation}

Where $\eta$ is the dynamic viscosity of the medium (water), $r$ is the radius of the particle and $\beta$ is a correction factor accounting for the proximity to the surface (wall effect). To calculate the  optical forces applied to particles by radiation pressure, the necessary values are are radius of micro particles - 512 nm, the refractive indices for water and glass, $n_{water}= 1.33$ and $n_{glass} = 1.52$ \cite{William2014}, respectively, and the refractive index for polystyrene at a wavelength of 1064 nm $n_{polyestyre}= 1.565$ \cite{Nikolov2000}. To determine the value of $E^2_0$, the relationship between the optical intensity, the magnitude of the electric field, and the area of incidence of the beam was used \cite{Griffiths2017}, which is given by

\begin{equation}
    I_{opt} = \frac{P_{measure}}{A_{region}}=\frac{1}{2}c\epsilon_0E^2_0
\end{equation}

Where $c=299.792.458$ (m/s) is speed of light, $\epsilon_0= 8.854\times 10^{-12}$ (F/m) is vacuum permittivity and $I_{opt}= 1.8$ in Watts (W) units (45$\%$ of the maximum current).

\begin{figure}[ht]
    \centering
    \includegraphics[width=\linewidth]{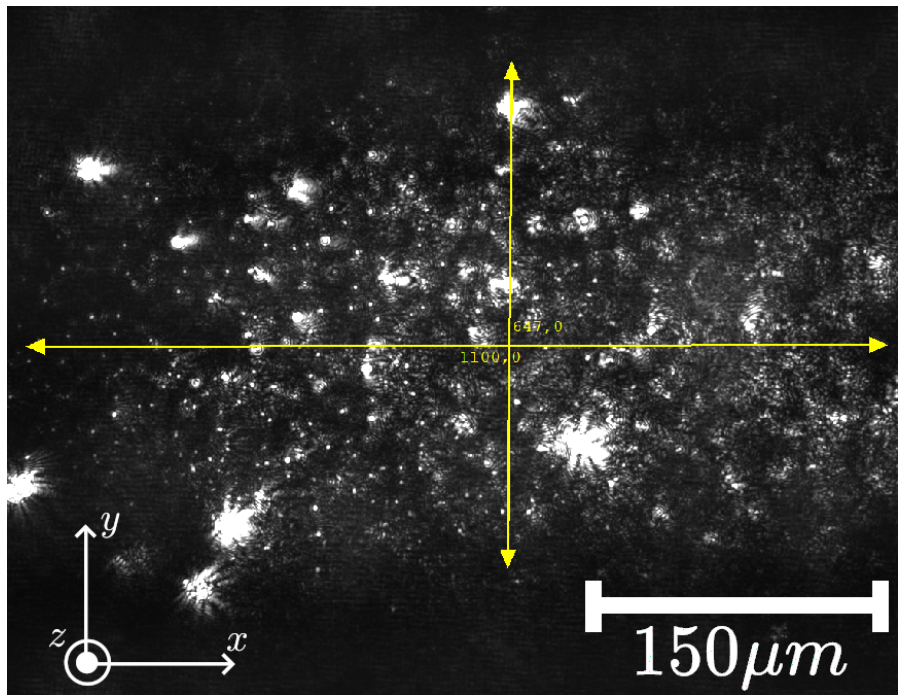}
      \caption{Measurement (in pixels) of laser beam area at prism surface area using formula of ellipse area $\pi ab$, where $a,b$ are major and minus axis, respectively. Software used was Thorcam installed with the CMOS monochrome sensor camera.}
        \label{fig6_EW.png}
    \end{figure}

To determine the beam's area of incidence, a lens with 20X magnification was used, allowing observation of the entire beam area. Then, using Thorcam software from Thorlabs\textsuperscript{\copyright}, $A_{region}$ was measured in pixels, as shown in Figure \ref{fig6_EW.png}. After calibrating the pixel-distance relationship for this lens, a beam area of $6.66 × 10^{-7}$m$^2$ was obtained.

The theoretical model predicts forces in the range of picoNewtons (pN), which corresponds to velocities in the order of 10--15 $\mu$m/s. Our experimental data shows high consistency with the theoretical values. The slight discrepancies are attributed to the exponential decay of the evanescent field and the sensitive dependence on the particle-surface distance ($z$). This alignment confirms that the experimental setup accurately captures the physics of evanescent wave interaction, providing a solid baseline for the subsequent analysis of biological samples (red blood cells), as the forces involved are well-characterized and predictable.

\begin{figure}[ht]
    \centering
    \includegraphics[width=\linewidth]{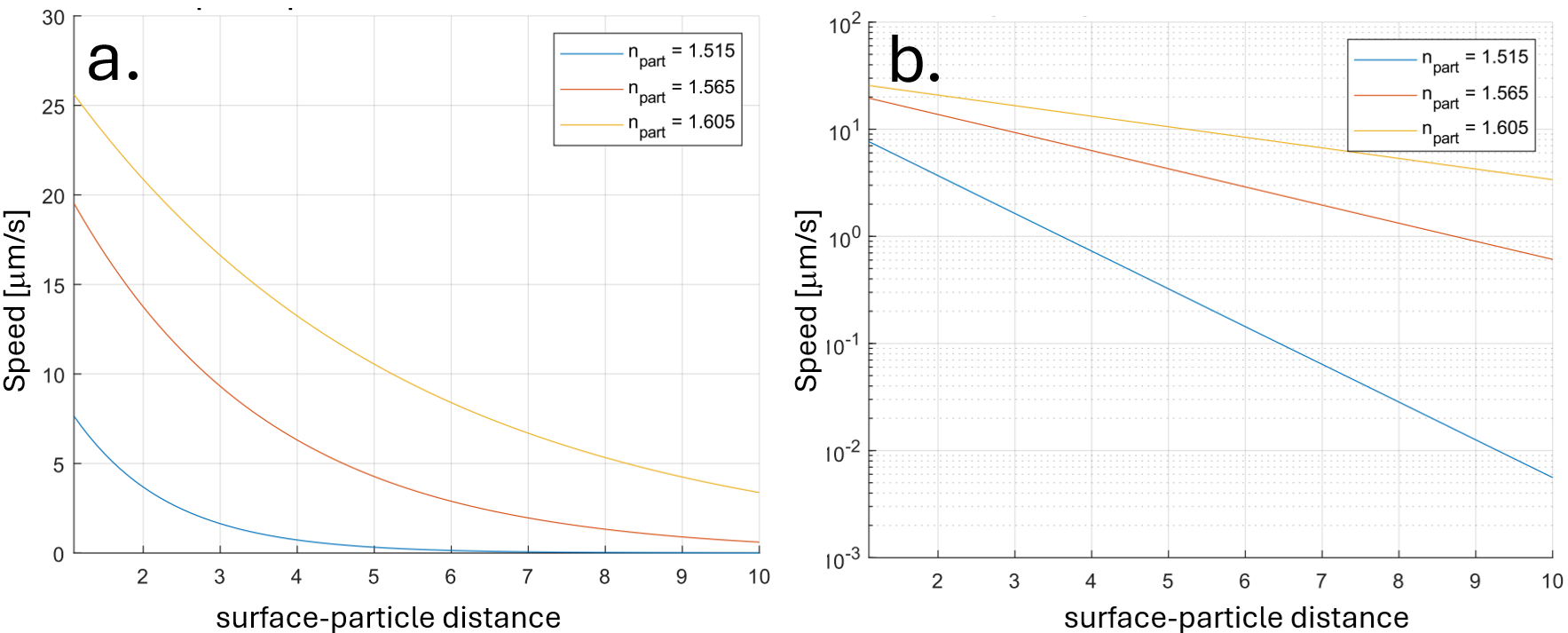}
    \caption{Predicted velocities of spherical micro-particles using $Matlab$ according to Stokes law $F_{drag}=\gamma v_{part}$ where $\gamma = 6\pi\eta R$. Assuming  equal forces, then $v_{particle}=F_{EW}/3\pi\mu D$, where $D$ is particle diameter  for different refractive indices of the microparticles. a. Linear scale. b. Logarithmic scale.}
    \end{figure}

\section{Results and discussion}

The quantification of particle and cell dynamics was performed using the TrackMate\textsuperscript{\copyright} plugin in Fiji/ImageJ which is widely used in the biological sciences, specifically for tracking cells, organelles, proteins, and other substances \cite{Trackmate2017}. To ensure consistency across measurements, the following digital signal processing parameters were established based on the physical dimensions of the subjects: (i) For polystyrene micro-particles ($D \approx 1 \mu m$), an estimated object diameter of 10 pixels was used as detection (Log Detector). For erythrocytes, the diameter was increased to 30 pixels to account for their larger cross-sectional area. (ii) The quality threshold was determined by identifying the first stable plateau in the quality histogram to filter out background noise while retaining real trajectories. (iii) A maximum linking distance of 25 pixels between consecutive frames was set Linking (Simple LAP Tracker parameter). (iv) To account for temporary signal loss (flickering), a maximum gap-closing distance of 10 frames was allowed. (v) Only trajectories containing at least 10 valid positions were included in the final velocity calculations to minimize the impact of Brownian noise on average speed estimation (Data Filtering).

\begin{figure}[ht]
    \centering
    \includegraphics[width=\linewidth]{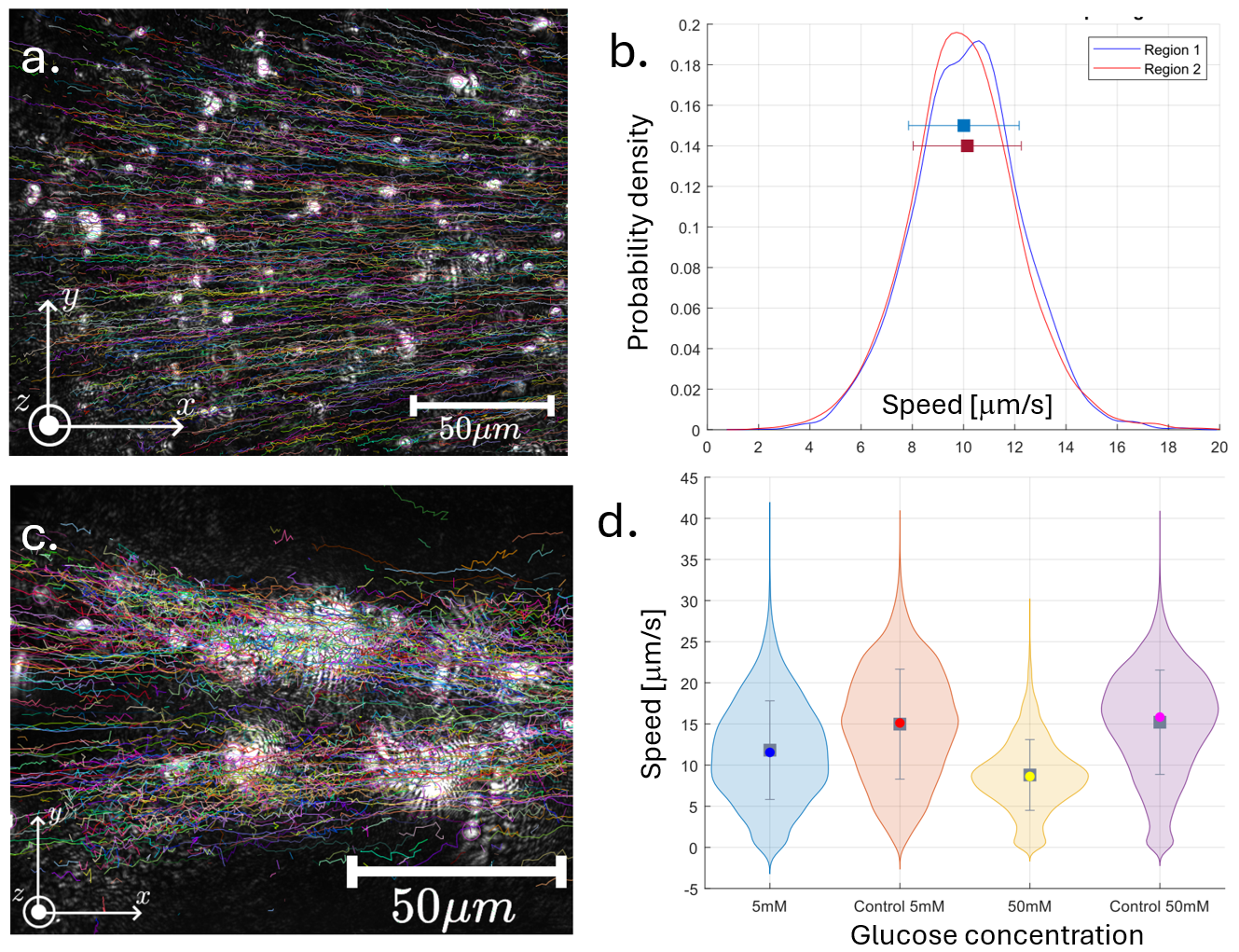}
       \caption{a. Representation of all paths traveled by the micro-particles generated in TrackMate\textsuperscript{\copyright}. b. Comparison of probability distributions for micro-particle velocities for two regions on the same prism (two-chamber method). c. Trajectories described by the erythrocytes, generated in $TrackMate$. d. Violin plot for red blood cell measurements at different glucose concentrations and their respective controls (with respect to micro-polystyrene particles); the gray bars correspond to the respective mean and standard deviation, while the colored dots indicate the medians of the data set.}
        \label{fig3_results}
    \end{figure}

With the above parameter settled, hundreds of individual trajectories were generated. The velocity distribution (Fig. \ref{fig3_results}) shows a characteristic peak for microparticles at 13 $\mu$m/s. A total of 64 data sets across 8 independent experiments were analyzed. The mean velocity for the 5 mM group was $11.8 \pm 2.1 \mu$m/s, while the 50 mM group showed a reduced mobility of $8.8 \pm 1.8 \mu$m/s.

\begin{table}[h]
\centering
\caption{Comparative Mobility Statistics}
\begin{tabular}{lccc}
\toprule
Sample & Glucose [mM] & Vel. [$\mu$m/s] & Std. Dev. \\
\midrule
RBC & 5 & 11.8 & 2.1 \\
RBC & 50 & 8.8 & 1.8 \\
Control & N/A & 15.2 & 1.4 \\
\bottomrule
\end{tabular}
\end{table}

The significant reduction in velocity ($p=0.019$) indicates that high glucose levels modify the cell's interaction with the evanescent field. This could be due to increased membrane rigidity (stiffening), alterations in the effective refractive index of the cell and/or changes in the cell-surface interaction area \cite{Tapia2021,Bravo2023}.

\begin{figure}[ht]
    \centering
    \includegraphics[width=\linewidth]{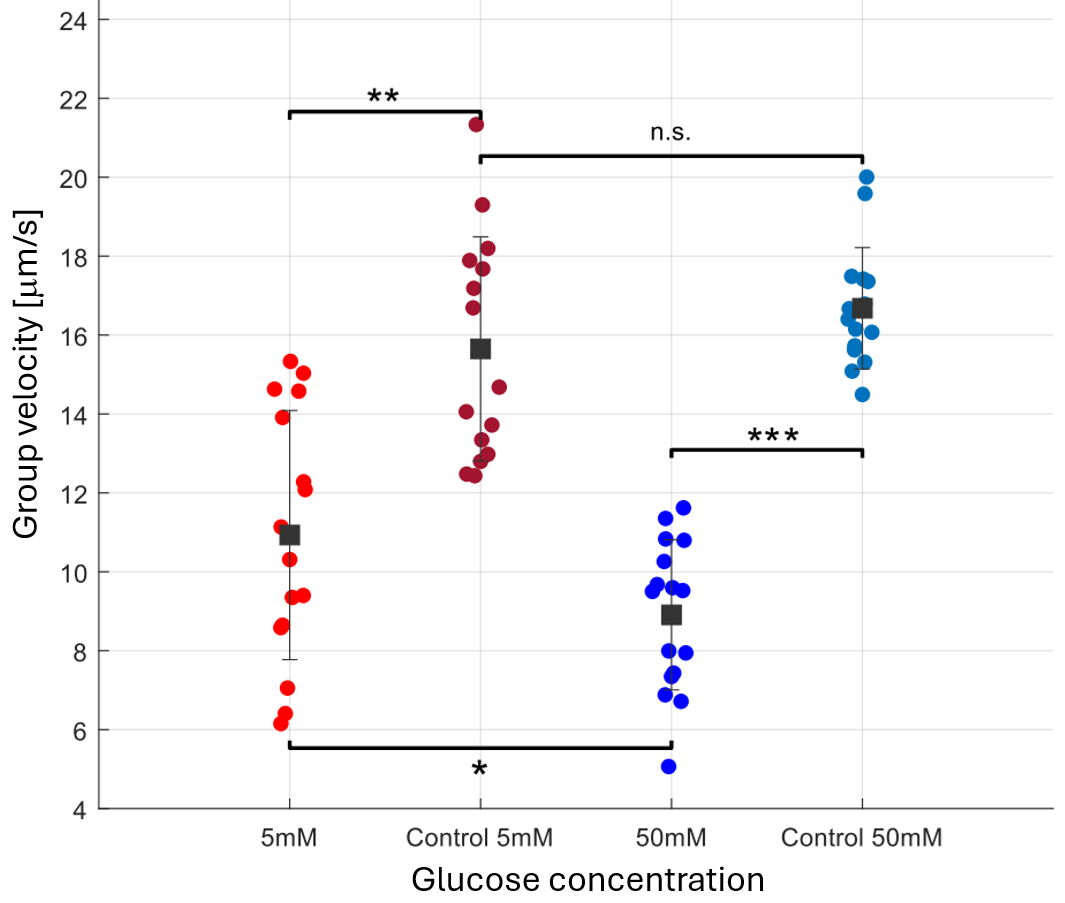}
    \caption{Average velocities for the different measurements performed on red blood cells at different glucose concentrations. Each group contains the average velocities of the 16 measurements performed for each case studied, with their mean and standard deviation represented by the gray line associated with each set. Statistically significant differences can also be observed between all data sets except for the two control samples.}
    \end{figure}

\section{Conclusions}

We have demonstrated a robust platform for the optical manipulation of biological cells using evanescent waves. The quantification of mobility differences under varying glucose concentrations provides a new metric for assessing cell health through micro-rheological responses. Despite these registers the system is still not fully upgraded. Although the experiments were conducted in a controlled environment, the lack of an active temperature control system may have introduced subtle changes in membrane elasticity. Furthermore, high concentrations of erythrocytes were observed to induce localized fluid flows, which contributed to the displacement of cells located beyond the theoretical penetration depth ($d_p \sim 250$ nm). Calibration using a USAF 1951 test target determined a spatial resolution of $0.163$ $\mu$m/pixel. The error associated with this calibration propagates to the final velocity distributions, particularly impacting the lower end of the measured speeds where pixel-to-pixel displacement is minimal.

Future experiment will incorporate these improvements as well interaction with randomly distributed non gaussian laser fields as speckle and bio-speckle for image pattern formation \cite{Staforelli2010,Lencina2012,Toderi2020}. This approach  could enable the development of  non-invasive optical tools for clinical diagnosis in vascular pathologies. Additionally, characterization of vegetal cells dynamics can also be incorporated to determine, in very particular, floral content in honeys and to identify adulteration. In this aspect, pollen group velocities can depend non only of water, glucose and pollen content but other types of syrups concentrations such as fructose, maltose or cane sugar \cite{Machuca2022,Jofre2025}. The particular blend can affect the dynamic viscosity of the sample medium $\eta$, the Stoke's drag forces and the refraction index \cite{Staforelli2026}. Finally, the use of neuromorphic vision based sensors are proposed. Traditional Particle imaging velocimetry relies on high-speed cameras to map time-resolved fluid flows (as the CMOS sensor used in this work), often leading to bandwidth and memory bottlenecks due to high resolutions and frame rates. Event-based cameras offer a solution to these limitations by recording only brightness changes, allowing for ultra-low latency and significantly reduced data rates \cite{Willert2022,Huenchual2025}.

\section*{Acknowledgments}
This work was supported by FONDEF IT24i0064, FONDECYT Iniciación 11230941 and the Facultad de Ciencias Físicas y Matemáticas, Universidad de Concepción (UDEC).


\begin{thebibliography}{22}
\bibitem{Petra2018} Petra Paiè, Tommaso Zandrini, Rebeca Martínez Vázquez, Roberto Osellame, and Francesca Bragheri. Particle manipulation by optical forces in microfluidic
devices. Micromachines, 2018. doi: https://doi.org/10.3390/mi9050200.
\bibitem{Ashkin2000} A. Ashkin. History of optical trapping and manipulation of small-neutral particle, atoms, and molecules. IEEE Journal of Selected Topics in Quantum
Electronics, 2000. doi: https://doi.org/10.1109/2944.902132.
\bibitem{Kishan2006}Kishan Dholakia and Peter Reece. Optical micromanipulation takes hold.
Nano Today, 2006. doi: https://doi.org/10.1016/S1748-0132(06)70019-6.
\bibitem{Volpe2017}Resnick, A. (2017). Optical tweezers: principles and applications, by Philip H. Jones, Onofrio M. Maragò, and Giovanni Volpe: Scope: manual, handbook, guide. Level: advanced undergraduate, postgraduate, early career researcher, teacher. 
\bibitem{Mark2015}Mark Daly, Marios Sergides, and Síle Nic Chormaic. Optical trapping and manipulation of micrometer and submicrometer particles. Laser \& Photonics Reviews, 2015. doi: https://doi.org/10.1002/lpor.201500006.
\bibitem{Milan2013}Milan Milosevic. On the nature of the evanescent wave. Applied Spectroscopy, 2013. doi: https://doi.org/10.1366/12-06707.
\bibitem{Oleg2020}Oleg V. Angelsky, Claudia Yu Zenkova, Steen G. Hanson, and Jun Zheng. Extraordinary manifestation of evanescent wave in biomedical application. Frontiers in Physics, 2020. doi: https://doi.org/10.3389/fphy.2020.00159.
\bibitem{Derek2001}Derek Toomre and Dietmar J. Manstein. Lighting up the cell surface with
evanescent wave microscopy. Trends in Cell Biology, 2001. doi: https://doi.org/10.1016/S0962-8924(01)02027-X.
\bibitem{Taitt2016}Chris Rowe Taitt, George P. Anderson, and Frances S. Ligler. Evanescent wave fluorescence biosensors: Advances of the last decade. Biosensors and Bioelectronics, 2016. doi: https://doi.org/10.1016/j.bios.2015.07.040.
\bibitem{Oleg2019}Oleg V. Angelsky, Peter P. Maksymyak, Claudia Y. Zenkova, Andrew P. Maksymyak, Steen G. Hanson, and Dimitrov D. Ivanskyi. Peculiarities of control of erythrocytes moving in an evanescent field. Journal of Biomedical Optics, 2019. doi: https://doi.org/10.1117/1.JBO.24.5.055002.
\bibitem{Oleg2019b}Oleg Angelsky, Claudia Zenkova, P.P. Maksymyak, A.P. Maksymyak, and
Dmytro Ivanskyi. Controlling and manipulation of red blood cells by evanescent waves. Optica Applicata, 2019. doi: 10.37190/oa190406.
\bibitem{Tapia2021}Tapia, J., Vera, N., Aguilar, J., González, M., Sánchez, S. A., Coelho, P., ... \& Staforelli, J. (2021). Correlated flickering of erythrocytes membrane observed with dual time resolved membrane fluctuation spectroscopy under different d-glucose concentrations. Scientific Reports, 11(1), 2429.
\bibitem{Bravo2023}Bravo, Nicolás, et al. "Flickering of fetal erythrocytes membrane under gestational diabetes observed with dual time resolved membrane fluctuation spectroscopy." Biochemistry and Biophysics Reports 36 (2023): 101556. 
\bibitem{Mario2017}Mario Bertolotti, Concita Sibilia, and Angela M. Guzman. Evanescent Waves in Optics. Springer Cham, 2017. ISBN 978-3-319-61261-4. doi: https://doi.org/10.1007/978-3-319-61261-4.
\bibitem{Griffiths2017}David J. Griffiths. Introduction to Electrodynamics. Cambridge University Press, 4th edition, 2017. ISBN 978-1108420419.
\bibitem{Nieto2011} Manuel Nieto-Vesperinas and J. Ricardo Arias-Gonzalez. Theory of forces
induced by evanescent fields. 2011. doi: https://doi.org/10.48550/arXiv.1102.1613.
\bibitem{Trackmate2017}Jean-Yves Tinevez, Nick Perry, Johannes Schindelin, Genevieve M. Hoopes, Gregory D. Reynolds, Emmanuel Laplantine, Sebastian Y. Bednarek, Spencer L. Shorte, and Kevin W. Eliceiri. Trackmate: An open and extensible platform for single-particle tracking. Methods, 2017. doi: https: //doi.org/10.1016/j.ymeth.2016.09.016.
\bibitem{Cho2018}Cho, N. H., Shaw, J. E., Karuranga, S., Huang, Y., da Rocha Fernandes, J. D., Ohlrogge, A. W., \& Malanda, B. I. D. F. (2018). IDF Diabetes Atlas: Global estimates of diabetes prevalence for 2017 and projections for 2045. Diabetes research and clinical practice, 138, 271-281.
\bibitem{Almaas1995}E. Almaas and I. Brevik. Radiation forces on a micrometer-sized sphere in an evanescent field. J. Opt. Soc. Am. B, 1995. doi: 10.1364/JOSAB.12.002429.
\bibitem{William2014}William M. Haynes, editor. CRC Handbook of Chemistry and Physics. CRC Press, 95th edition, 2014. ISBN 978-1482208689.
\bibitem{Nikolov2000}Ivan D. Nikolov and Christo D. Ivanov. Optical plastic refractive measurements in the visible and the near-infrared regions. Appl. Opt., 2000.doi: 10.1364/AO.39.002067.
\bibitem{Staforelli2010}Staforelli, J. P., Brito, J. M., Vera, E., Solano, P., \& Lencina, A. (2010). A clustered speckle approach to optical trapping. Optics communications, 283(23), 4722-4726.
\bibitem{Lencina2012}Lencina, A., Solano, P., Staforelli, J. P., Brito, J. M., Tebaldi, M., \& Bolognini, N. (2012). Three-dimensional clustered speckle fields: theory, simulations and experimental verification. Optics Express, 20(19), 21145-21159.
\bibitem{Toderi2020}Toderi, Martín A., Bibiana D. Riquelme, and Gustavo E. Galizzi. "An experimental approach to study the red blood cell dynamics in a capillary tube by biospeckle laser." Optics and Lasers in Engineering 127 (2020): 105943.
\bibitem{Machuca2022}Machuca, G., Staforelli, J., Rondanelli-Reyes, M., Garces, R., Contreras-Trigo, B., Tapia, J., ... \& Coelho, P. (2022). Hyperspectral microscopy technology to detect syrups adulteration of endemic guindo santo and quillay honey using machine-learning tools. Foods, 11(23), 3868.
\bibitem{Jofre2025}Jofre, R., Tapia, J., Troncoso, J., Staforelli, J., Sanhueza, I., Jara, A., ... \& Coelho, P. (2025). YOLOv8-based on-the-fly classifier system for pollen analysis of Guindo Santo (Eucryphia glutinosa) honey and assessment of its monoflorality. Journal of Agriculture and Food Research, 19, 101665.
\bibitem{Staforelli2026}Staforelli-Vivanco, J., Salamanca-Levi, V., Jofré-Cerda, R., Rondanelli-Reyes, M., \& Lamas, I. (2026). Three-Dimensional Volumetric Reconstruction of Native Chilean Pollen via Lens-Free Digital In-line Holographic Microscopy. arXiv preprint arXiv:2601.14205.
\bibitem{Willert2022}Willert, C. E., \& Klinner, J. (2022). Event-based imaging velocimetry: an assessment of event-based cameras for the measurement of fluid flows. Experiments in Fluids, 63(6), 101.
\bibitem{Huenchual2025}Huenchual-Escobar, J., Solano, P., Staforelli, J., Vera, E., \& Tapia, J. (2025). Boosting microparticle tracking with neuromorphic cameras by optical modulation. Scientific Reports, 15(1), 35299.




\end{thebibliography}
\end{document}